\documentclass[
twocolumn,
reprint,
superscriptaddress,
frontmatterverbose, 
nofootinbib,
 amsmath,amssymb,
floatfix,
tightenlines,
]{revtex4-1}
\usepackage{dcolumn}
\usepackage{bm}
\usepackage[pdftex]{graphicx}
\usepackage{amssymb,amsfonts,amsmath}
\usepackage{color}
\usepackage[usenames,dvipsnames]{xcolor}

\begin{document}
\title{Transition in heterogeneous dynamics across the morphological hierarchy in two-dimensional aggregates}
\author{Tamoghna Das}
\email[electronic address: ]{tamoghna.das@oist.jp}
\affiliation{Collective Interactions Unit, Okinawa Institute of Science and Technology Graduate University, Onna, Okinawa 9040495, Japan}
\author{T. Lookman}
\email[electronic address: ]{txl@lanl.gov}
\affiliation{Theoretical Division, Los Alamos National Laboratory, Los Alamos, New Mexico 87545, USA}
\author{M. M. Bandi}
\email[corresponding author: ]{bandi@oist.jp}
\affiliation{Collective Interactions Unit,  Okinawa Institute of Science and Technology Graduate University, Onna, Okinawa 9040495, Japan}
\date{\today}

\begin{abstract}
Two-dimensional (2D) particulate aggregates formed due to competing interactions exhibit a range of non-equilibrium steady state morphologies from finite-size compact crystalline structures to non-compact string-like conformations. We report a transition in heterogeneous microscopic dynamics across this morphological hierarchy as a function of decreasing long-range repulsion relative to short-range attraction at a constant {\it low} density and temperature. Following a very slow cooling protocol to form steady state aggregates, we show that geometric frustration inherent to competing interactions assures non-ergodicity of the system, which in turn results in long-time sub diffusive relaxation of the same. Analysing individual particle trajectories generated by molecular dynamics, we identify {\it caging} dynamics of particles in compact clusters in contrast to the {\it bonding} scenario for non-compact ones. Finally, by monitoring temperature dependence, we present a generic relation between diffusivity and structural randomness of the aggregates, irrespective of their thermodynamic equilibrium.

\end{abstract}
\maketitle

\section{Introduction}
Relating the morphology and dynamics of non-equilibrium systems has remained a challenge despite concerted theoretical and experimental efforts \cite{effort1,effort2,effort3,effort4} over the years. Systems driven out of equilibrium, such as glassy and granular materials, exhibit dynamic structural order \cite{DH1, DH2} which often extends over different length and time scales without any clear separation between them. The effect of geometry on dynamics \cite{conf1, conf2} is also expected in systems confined by external fields or porous media. Transport within crowded environments such as living cells \cite{cell} presents yet another example where spatial structure and dynamics are surrogate to each other. Here, we present a detailed study of both global and local dynamics of a generic non-equilibrium pattern forming particulate system, namely, 2D aggregates. Aggregates of particles formed due to competing interactions enjoy a broad range of structural randomness (or order) \cite{aggrgt1, aggrgt2, aggrgt3} intermediate between liquids and crystalline solids. By following the steady state trajectories simulated by molecular dynamics, we establish a phenomenological correspondence between the local morphology and dynamics of 2D aggregates.

Experimentally, the aggregates are usually formed by quenching a high-temperature equilibrium liquid \cite{expt1, expt2} to a much lower temperature. While such a dynamical preparation protocol is useful in practice, identification of the governing dynamics and a theoretical understanding of the processes becomes difficult. Different mechanisms such as diffusion-limited \cite{dbt1} and reaction-limited aggregation \cite{dbt2} have been reported. To describe the phase behaviour of such systems, both equilibrium routes \cite{dbt3, dbt4} and kinetic viewpoint \cite{dbt5, dbt6, dbt7} have been thoroughly explored. Stemming from a fast quench, the glass transition scenario has also been conjectured and quite thoroughly investigated \cite{glss1, glss2}. 

In the present work, a very slow cooling protocol is adopted for aggregate formation and the final temperature is maintained over a long observation time upon reaching the desired value. By monitoring the evolution of energy fluctuations during cooling, we show that frustration innate to competing interactions is the key factor for the non-equilibrium dynamics of aggregates. The steady state dynamics is non-ergodic, sub-diffusive and heterogeneous for all aggregate morphologies. While phase separation affects this overall dynamics, local slow dynamics results from the reduced thermal fluctuations at final `low' temperatures. We find that individual particle trajectories and their spatio-temporal  correlation reveal a continuous transition in microscopic dynamics in concert with the changing shape of aggregates. Finally, we demonstrate the scaling of the average dynamics specified by scaled diffusivity of aggregates with respect to their local geometry quantified by the excess entropy. This relation points to a fundamental connection between average dynamics and structural randomness of a system irrespective of its thermal equilibrium.

The organisation of the paper is as follows: we introduce the specific form of competing interactions used in the study in Sec.II and briefly describe the variations of morphologies obtained by tuning the competition. The details of cooling protocol and molecular dynamics simulation are charted out in Sec.III. In Sec.IV, we elaborate the main results on dynamics in three parts, (A) non-ergodicity, (B) global dynamics and (C) particle dynamics. A discussion and further implications of these results are presented in Sec.V followed by a brief summary.

\section{Variations in morphology from competing interactions}
Competing interactions are necessary for aggregate formation \cite{lebo_pen,compint} and can occur in diverse physical settings \cite{colloid1,protein1,protein2,dna_np}, each having its own set of control parameters \cite{salt}. Our choice of model system is a reliable representative of globular proteins \cite{globprot} and colloidal systems \cite{colloid2} with suitable polymer coating.  In this model, the short-range attraction and long-range repulsion are realised by a generalised Lennard-Jones potential \cite{2n-n} and a pair-wise potential of Yukawa form, respectively, 
\begin{eqnarray}
\phi_{SA} &=&4\epsilon[(\sigma/r)^{2\alpha}\!-\!(\sigma/r)^{\alpha}] \\
\phi_{LR} &=&(A\sigma/r)\exp(r/\xi)
\end{eqnarray}
Setting length and energy scales by $\sigma$ and $\epsilon$ respectively, the range of attraction is controlled by $\alpha$. The attraction range is $0.2\sigma$ for our choice $\alpha\!=\!18$. The same extends to $2.5\sigma$ for the usual Lennard-Jones potential with $\alpha\!=\!6$. We note that the thermodynamic properties of a particulate system with bare attraction is independent of $\alpha$ for $\alpha \ge 18$ \cite{2n-n}. The Yukawa part implicitly models the long-range repulsive effect of surrounding media. The strength of repulsion $A$ and the screening length $\xi$ are measured in $\epsilon$ and $\sigma$ units respectively. The effective centro-symmetric interaction, $\phi\!=\!\phi_{SA}\!+\!\phi_{LR}$, has a positive energy barrier at finite distance separated from the steep hard core repulsion by an attractive minimum. The height and extent of the barrier can be controlled by $A$ and $\xi$. In a previous study \cite{2Dmorph}, we have shown that this system exhibits a structural hierarchy at a constant density and temperature for different realisations of competition controlled by $\alpha, A$ and $\xi$. Aggregate morphology shows a continuous transition from finite-size non-compact to compact to percolated {\it gel} structures as a function of decreasing repulsion against attraction.

\begin{figure}[h!]
\begin{center}
\includegraphics[width=0.9\linewidth]{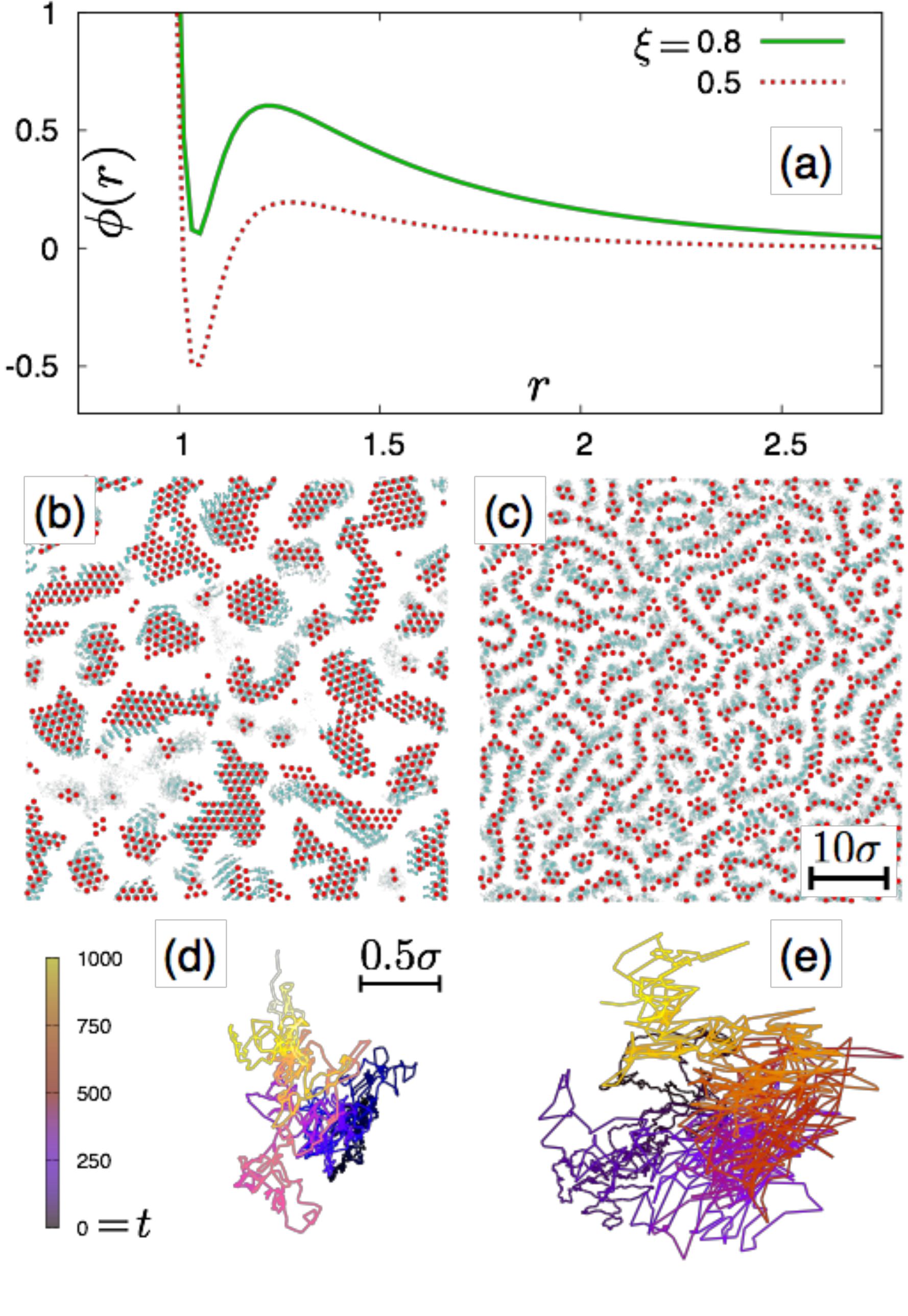}
\end{center}
\caption{
(color online) (a) The form of effective interaction $\phi(r)$ is drawn for two different values of $\xi\!=\!0.5$ and $0.8$ seting $A\!=\!4.0$ and $\alpha\!=\!18$. Please note that the global attractive minimum becomes local with increasing $\xi$. Visual representation of particle dynamics for these two $\xi$ values are shown in (b) compact ($\xi\!=\!0.5$) and (c) non-compact ($\xi\!=\!0.8$) structures, respectively, at $T\!=\!0.05$. Only a part of the simulation box is shown for clarity. The scale used for both (b) \& (c) is same and shown in (c). Each {\it red} filled circle denotes an instantaneous particle position. {\it Blue} background consists of dots representing the positions of the same particles for $100$ earlier instants stored in a regular interval of $500$ simulation steps. (e) \& (f) are trajectories of a single particle belonging to (c) \& (d) respectively, colour coded by time. Using the same scale (scale bar $\!=\!0.5\sigma$), it becomes evident that the particle trajectory (d) in compact clusters is more restricted and jagged than that in (e) non-compact clusters.
}
\label{pot}
\end{figure}
In this study, we focus on a restricted parameter space of only finite-size clusters and investigate the dynamics across compact to non-compact morphologies. Specifically, we present the detailed analysis of microscopic dynamics for two representative cases, $\xi\!=\!0.5$ and $0.8$, showing compact and non-compact aggregates respectively for fixed $A\!=\!4.0$. We mention that the strength of repulsion equals that of attraction for this choice of $A$. The form of the effective pair-wise interaction $\phi$ is shown in Fig.\ref{pot}(a). Note the dominant attraction well for $\xi\!=\!0.5$ which is complemented by a small repulsive barrier slowly decaying at large $r$. The resulting configuration consists of perfect crystalline islands of exponentially distributed size. Percolated structures appear for lower values of $\xi$ which we keep out of the present study. When $\xi$ is raised to $0.8$ keeping $A$ constant, the repulsive barrier increases significantly and the attractive minimum close to the particle core ($r\!=\!0$) becomes comparable with the repulsive minimum far from the core. Finite-size one-dimensional arrangements of particles are observed under such strong influence of repulsion. Upon further increase of $\xi$, the size of non-compact chain-like aggregates becomes smaller and the overall randomness in the configuration increases further.

A part of the simulation box is presented in Fig.\ref{pot}(b)\&(c) showing both of the limiting morphologies, compact and non-compact, respectively. {\it Red} filled circles in these figures represent instantaneous positions of particles. Positions of the same particles for $100$ previous instants are plotted in the background with {\it blue} dots. All the collated configurations are separated from each other by $500$ molecular dynamics steps. The slow dynamics experienced by the particles is already evident from these visual representations. Another important observation is the consistency of collective dynamics of particles. Particles forming a cluster undergo collective translation and/or rotation; even bending to change conformations. However, fragmentation and reformation of aggregates are seldom observed. In other words, the system has reached its steady state at least in terms of the size of clusters. Trajectories of a single particle chosen arbitrarily from the compact and non-compact clusters are drawn in Fig.\ref{pot}(d)\&(e) respectively with colours according to their time evolution. The spatial extent of the trajectory is restricted for compact aggregates compared to that of a non-compact one. The former appears to be more heterogeneous than the latter from visual inspection. We analyse these trajectories in terms of several measures of correlation to understand the microscopic dynamics of the systems. Before discussing those results, we briefly describe the cooling protocol observed in this study and comment on the energy fluctuation during and after cooling in the next section. 

\section{Cooling protocol}
\begin{figure}[h!]
\begin{center}
\includegraphics[width=0.85\linewidth]{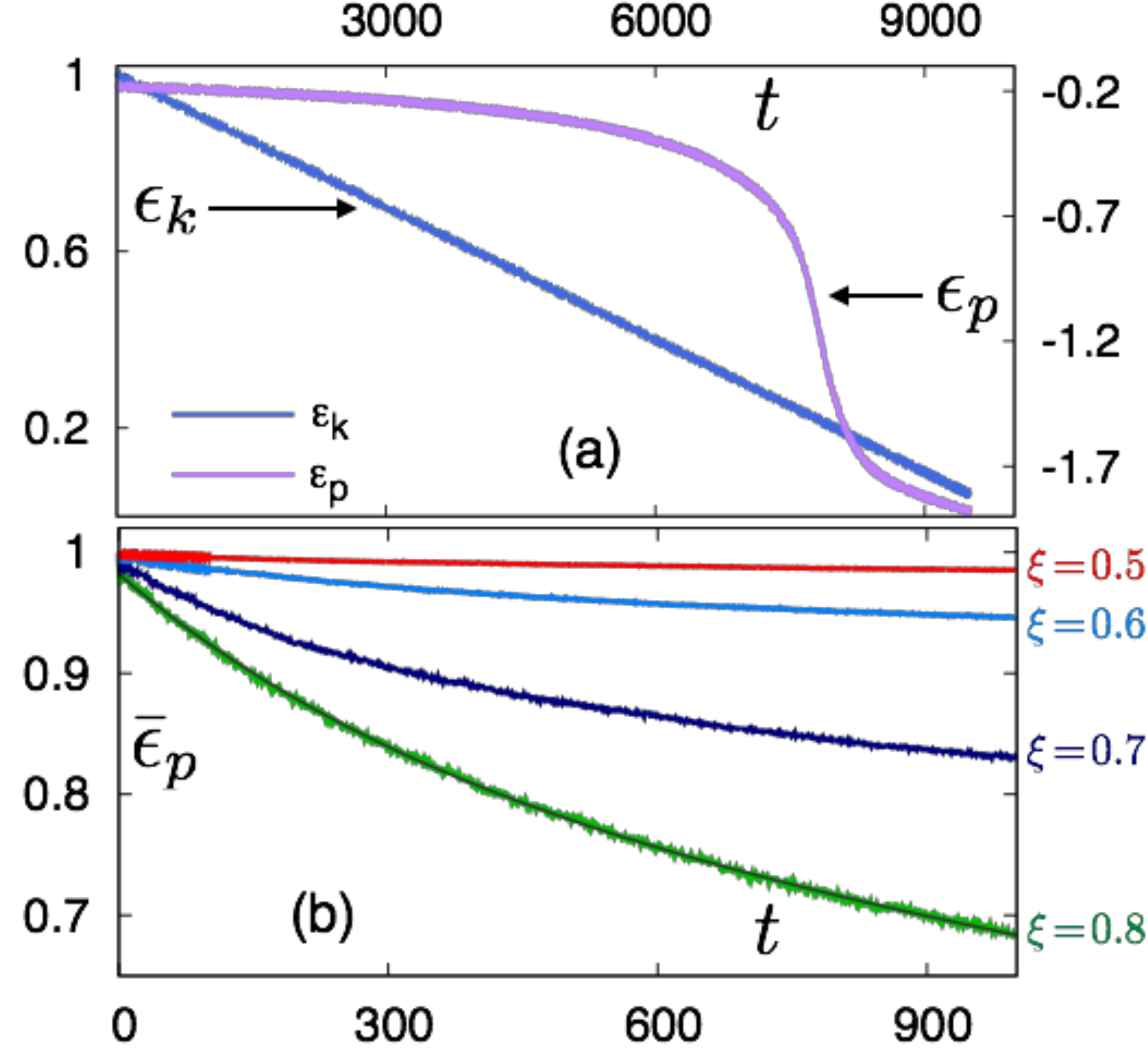}
\end{center}
\caption{
(color online) (a) Linear cooling denoted by decreasing average instantaneous kinetic energy $\epsilon_k$ ({\it solid blue line}) and corresponding variation of average potential energy $\epsilon_p$ ({\it solid purple line}) are plotted. Please note the sharp decrease in $\epsilon_p$ near $\epsilon_k\!=\!0.2$ which is close to the gas-liquid critical temperature for a system with bare attractive potential. (b) Non-exponential decay of normalised $\epsilon_p$ ({\it see text}) at fixed final temperature $T\!=\!0.05$ is shown for different values of $\xi\!=\!0.5$ ({\it red}), $0.6$ ({\it light blue}), $0.7$ ({\it dark blue}) and $0.8$ ({\it green}). 
}
\label{cool}
\end{figure}

We consider a $2D$ system of interacting mono-disperse particles at fixed density $\rho\!=\!0.4$. The initial configuration is a random non-overlapping arrangement of $N\!=\!56000$ particles of unit mass in a $376.0\sigma\!\times\!372.0\sigma$ box with periodic boundary conditions along all directions. Particle trajectories in canonical ensemble are generated by following molecular dynamics (MD) at constant number-area-temperature ({\it NAT}) as implemented in {\it LAMMPS} simulation package \cite{lammps}. Temperature is measured in $\epsilon$ units with the Boltzmann constant, $k_B\!=\!1$ and maintained by a Langevin thermostat \cite{langevin}. While the particles experience forces from others due to pair-wise interaction $\phi$, they are also subject to damping from implicit surroundings and a random force of Gaussian variance. The motion of $i$-th particle with position vector ${\bf r}_i$ can then be expressed as follows,
\begin{equation}
\ddot{{\bf r}}_i = -\sum_{j\ne i} \nabla \phi(r)-\nu\dot{{\bf r}}_i +{\bf \zeta}_i
\end{equation}
considering the force due to interaction $\phi(r)$ and frictional drag $\nu\dot{{\bf r}}_i $ from implicit media experienced by the particle. ${\bf \zeta}_i$ is a random force with zero mean and Gaussian variance, $\langle \zeta_i(t_0)\zeta_j(t+t_0)\rangle = 2k_B (T/\nu)\delta_{i,j}\delta(t)$. Numerical integration is performed by using velocity Verlet algorithm with time steps $\delta t\!=\!10^{-3}\tau$ where $\tau\!=\!\sigma/\sqrt{\epsilon}$ is the unit of time for unit mass. The system is first equilibrated at temperature $T_i\!=\!1.0$ to form a simple liquid. We mention that the system goes from ballistic to diffusive regime in time slightly longer than $\tau$ and this feature holds over a range of temperature up to $T\!\sim\!0.2$. The temperature of the equilibrated liquid is then linearly ramped down over $9.5\!\times\!10^7$ MD steps to final temperature $T_f=0.05$ (Fig.\ref{cool}(a)). The temperature of the system thus drops by only $10^{-4}\epsilon$ over unit time $\tau$.

Instantaneous temperature or kinetic energy per particle $\epsilon_k$ is shown in Fig.\ref{cool}(a) during linear cooling. The potential energy per particle $\epsilon_p$, following $\epsilon_k$, shows an initial linear decay but drops abruptly around $T\!\sim\!0.2$. We note that the critical temperature of the bare attractive potential is $T_c\!=\!0.18\pm0.01$ \cite{critical}. This sharp drop of $\epsilon_p$ close to $T_c$ is indicative of the effect of phase separation process on the system. After reaching a desired final temperature, it is maintained by the same thermostat over a period of $10^3\tau$. This is our observation period when the trajectories are saved for further analysis. Most importantly, we note that the $\epsilon_p$ continues to decay and does not reach a steady state value over these three decades of time. This decay can be fitted well with a stretched exponential or a double-exponential function. In Fig.\ref{cool}(b), we show this non-exponential decay for different values of $\xi$. We have plotted the scaled potential energy, $\bar{\epsilon}_p\!=\!\epsilon_p(t)/\epsilon_p(0)$ to compare its evolution for different $\xi$. Without getting into the details of the fitting parameters, we would like to emphasise that our model system, although cooled very slowly, remains out of equilibrium over a considerable period of time.
 
\section{Results}
\subsection{Non-ergodicity}
\begin{figure}[b!]
\begin{center}
\includegraphics[width=0.85\linewidth]{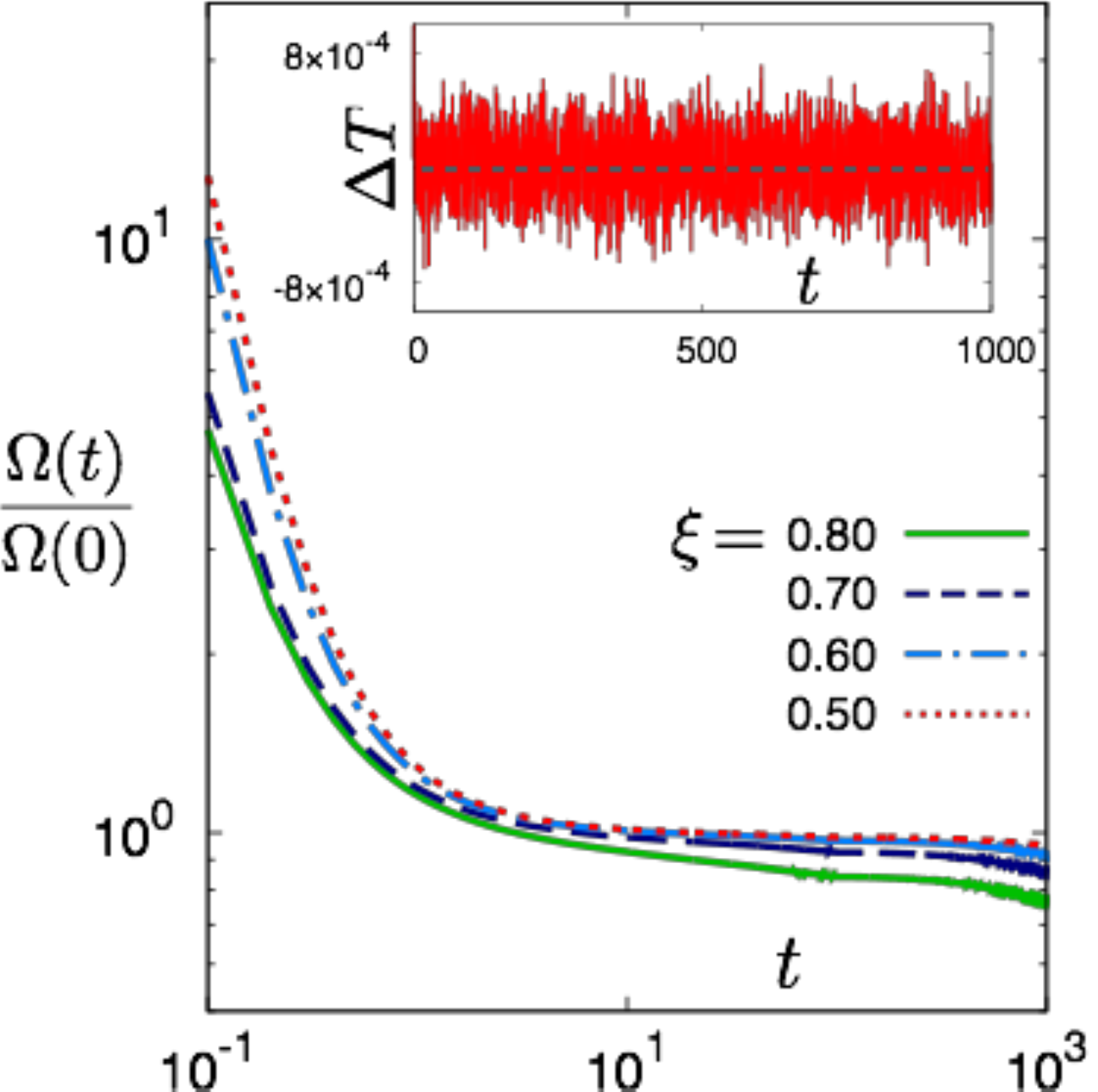}
\end{center}
\caption{
(color online)  The temperature fluctuation $\Delta T$ after cooling is shown in the {\it inset}. For a system approaching equilibrium, the normalised energy metric decays linearly with inverse of time ({\it see eq.\ref{emetric}}).  However, the long time decay is evidently much slower and thus the non-ergodicity of the aggregate systems over the observation time is established.
}
\label{nergo}
\end{figure}
Unusual time evolution of potential energy at constant $T$ prompts us to study its fluctuations further. We compute the {\it energy fluctuation metric} \cite{nergo1,nergo2} defined as follows:
\begin{equation}
\Omega_E(t) = 1/N\sum_i^N (\overline{e_i(t)}-\langle e(t)\rangle)^2
\end{equation}
for the ensemble of simulated particle trajectories with average energy $E$ over time $t$. For the $i$-th particle, its average energy upto time $t$, $\overline{e_i(t)}$ and ensemble averaged energy $\langle e(t)\rangle$ read as,
\begin{eqnarray}
\overline{e_i(t)}&=&1/t\int_0^tdt^\prime E_i(t^\prime) \\
\langle e(t)\rangle&=&1/N\sum_{i=1}^Ne_i(t)
\end{eqnarray}
According to the ergodic hypothesis,
\begin{equation}
\overline{e_i(t\!\to\!\infty)}\!=\!\langle e(t)\rangle\!=\!\langle E\rangle
\end{equation}
Hence, an equilibrium system should fulfil the {\it ergodic convergence criterion}, $\lim_{t\!\to\!\infty} \Omega_E(t)\!\to\! 0$. Following this, at large $t$, the energy fluctuation metric should behave as \cite{nergo2}
\begin{equation}
\frac{\Omega_E(t)}{\Omega_E(0)}\approx\frac{1}{D_Et}
\label{emetric}
\end{equation}
for a system approaching equilibrium. The energy diffusion constant $D_E$ is the rate at which the particles access the available conformational space. However, our observation clearly deviates from this (Fig.\ref{nergo}). The evolution of $\Omega(t)/\Omega(0)$ is much slower at long time. Temperature remains steady with nominal fluctuations (Fig.\ref{nergo}({\it inset})) over this whole period of observation.With this firm evidence of non-ergodicity of the model system prepared by slow cooling and uninfluenced by any other external field, we attribute this feature entirely to the inherent geometric frustration of competing interactions. 

\subsection{Average dynamics: mean square displacement and self-overlap statistics}
\begin{figure}[h!]
\begin{center}
\includegraphics[width=0.9\linewidth]{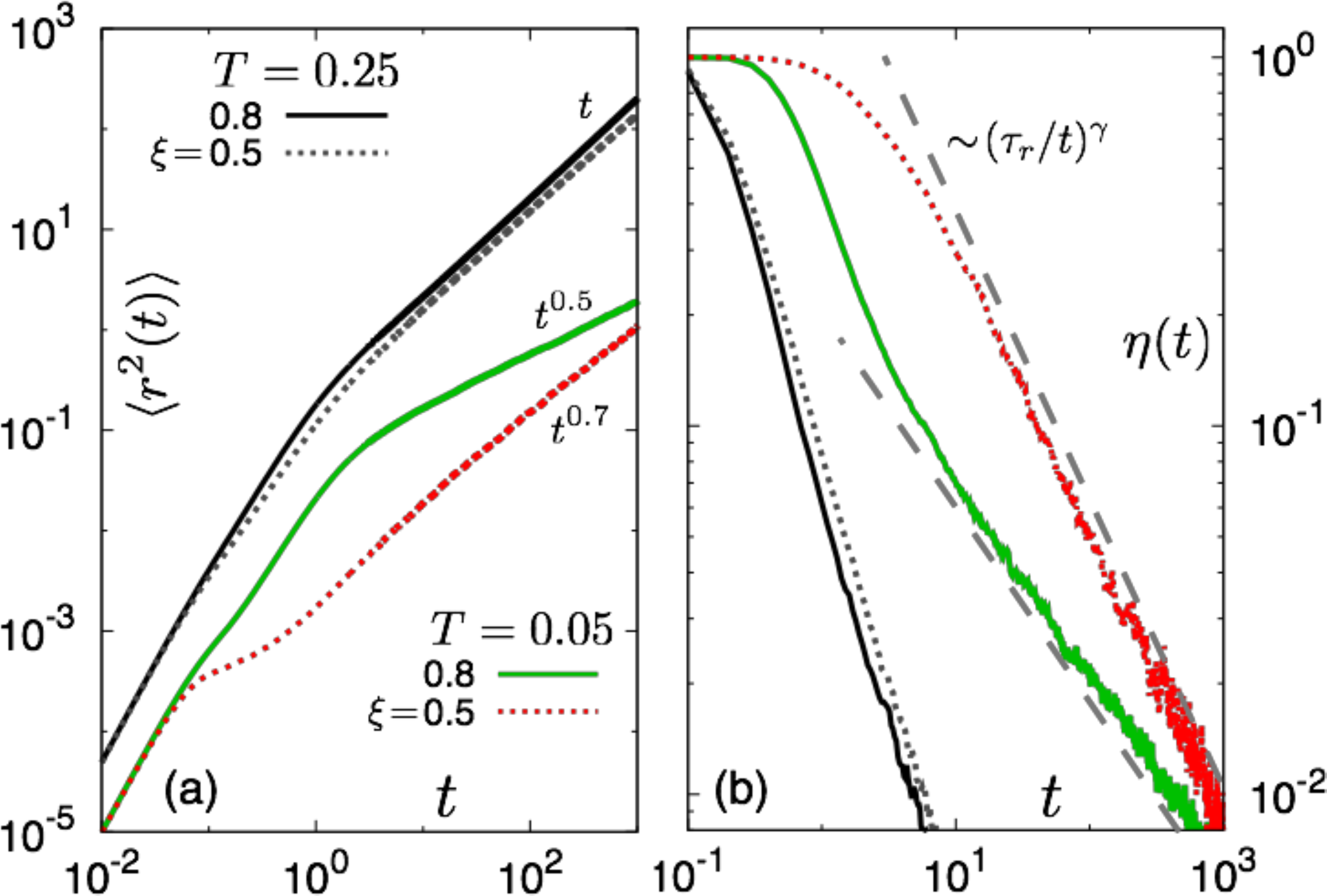}
\end{center}
\caption{
(color online) (a) Mean square displacement ({\it MSD}) of particles is plotted against time for compact ($\xi\!=\!0.5$) and non-compact ($\xi\!=\!0.8$) aggregates with fixed $A\!=\!0.4$. At low temperature, $T\!=\!0.05$, {\it MSD} shows three distinct regime, short-time ballistic, intermediate slow relaxation and long time sub-diffusive behaviour ($\langle r^2(t)\rangle\!\sim\!t^n$, $n\!<\!1$). Intermediate slow relaxation is more prominent for compact clusters than the non-compact ones. Both systems behave as a simple liquid at high temperature $T\!=\!0.25$ showing only short-time ballistic and long-time diffusive behaviour. (b) At low temperature ($T\!=\!0.05$), the self-overlap function $\eta(t)$ ({\it see text}) follows a power-law relation with time after an initial residence time $\tau_r$ which is larger for compact structures compared to the same for non-compact ones. Upon increasing temperature, it behaves like a normal equilibrium liquid (at $T\!=\!0.25$ and beyond).
}
\label{avdyn}
\end{figure}
\begin{figure*}[t!]
\begin{center}
\includegraphics[width=0.8\linewidth]{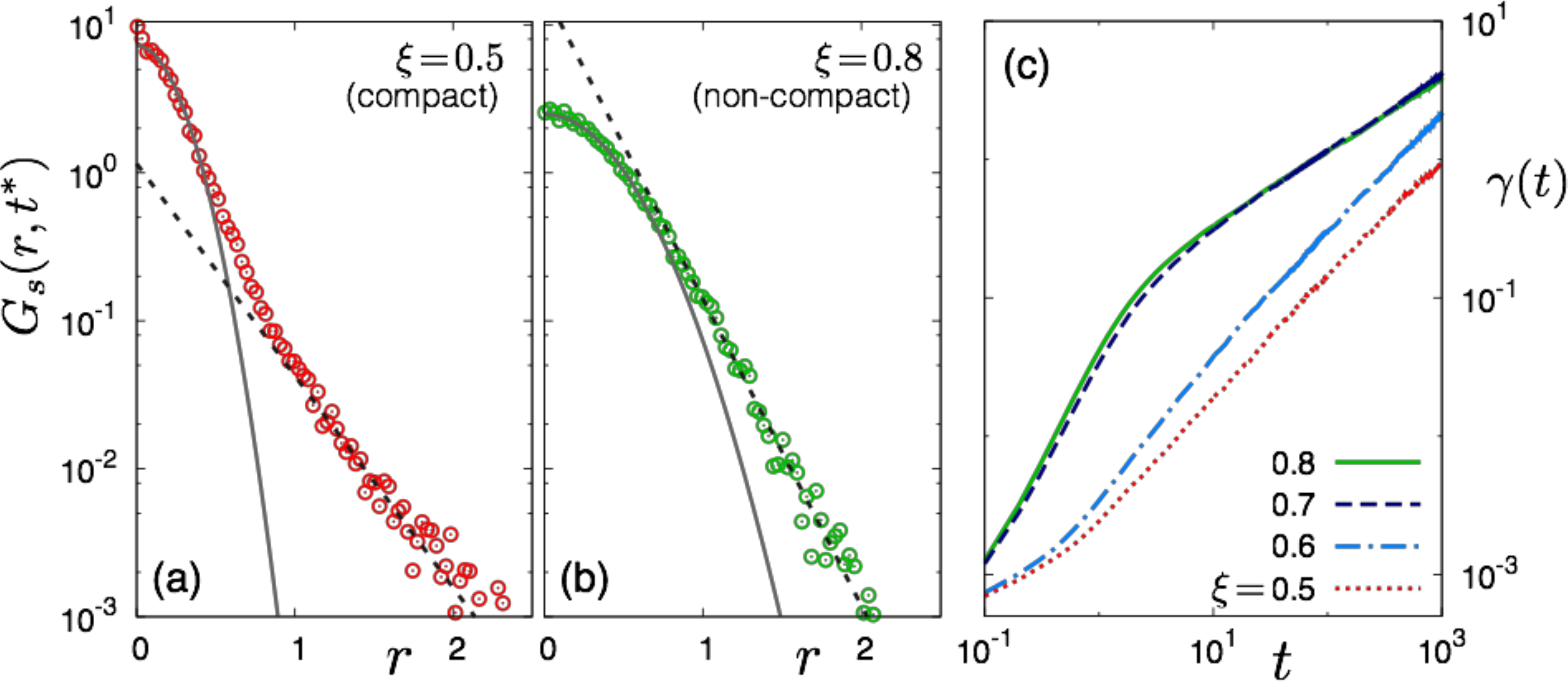}
\end{center}
\caption{
(color online) Non-Gaussian behaviour of self-displacement fluctuations $G_s(r,t^*)$ ($t^*\!=\!50\tau$) is shown for compact and non-compact aggregates in (a) and (b) respectively. Please note the change in microscopic dynamics accompanying the structural transition. While dominant Gaussian self-fluctuations end with an exponential tail for compact clusters, displacement of a particle from its arbitrary origin is distributed exponentially over space for non-compact morphologies. (c) Temporal evolution of dynamic Lindemann parameter $\gamma(t)$ measuring the relative displacement fluctuation of particles with respect to its neighbours captures another aspect of this dynamical transition. Very slow initial increase in $\gamma(t)$ for compact aggregates clearly points to the {\it caging} of particles by their near-neighbours. On the other hand, initial steep increase followed by significant slowing down of $\gamma(t)$ at long-time can be attributed to {\it bonding} in non-compact structures.
}
\label{lkldyn}
\end{figure*}
We now focus on analysing the dynamics of constituent particles in this non-equilibrium setting. We start with the mean square displacement of particles over the observation time. Mean square displacement ({\it MSD}) defined as,
\begin{equation}
\langle r^2(t)\rangle = 1/N \sum_{i=1}^N \{r_i(t)-r_i(0)\}^2
\end{equation}
measures the spatial extent of a particle's displacement from its origin after which the particle diffuses randomly due to thermal fluctuations. Time dependence of {\it MSD} is generically expressed as $\langle r^2(t)\rangle\!\sim\!t^n$ \cite{msd} where free diffusion is observed for $n\!=\!1$ and $n\!<\!1$ indicates sub-diffusion. As long as, the final temperature of our model system stays above $T\!=\!0.2$, long-time diffusive behaviour is observed (Fig.\ref{avdyn}(a)). Sub-diffusion at long-time appears for $T\!<\!0.2$. We show the {\it MSD} for both compact ($\xi=0.5$) and non-compact ($\xi=0.8$) structures at the lowest temperature $T=0.05$ reached by cooling in our simulations. Though sub-diffusion is evident for both types of structures, the exponent $n$ varies. Interestingly, $n$ for non-compact string-like structures is the same as the exponent for single-file diffusion \cite{sfd}. One important feature revealed from the {\it MSD} data is the prominent intermediate slow relaxation for compact morphologies. This feature is also observed for non-compact ones, though with less prominence. This intermediate relaxation is indicative of heterogeneous dynamics which we shall probe in the next section.

Here, we continue investigating average dynamics further to study the overlap between two arbitrarily chosen instantaneous configurations to probe the degree of structural relaxation within the system. Quantification of the overlap \cite{ovrlp1,ovrlp2} is computed as follows:
\begin{equation}
\eta(t)=\sum_{i,j=1}^N w(|{\bf r}_i(t)-{\bf r}_j(0)|)
\end{equation}
where $i,j$ are particle indices and $w(r)$ is a window function which is unity if $|{\bf r}_1\!-\!{\bf r}_2|\!<\!a$ and zero otherwise. Allowing a particle to relax within certain threshold distance $a$ from its original position, this quantity tells us whether a particle is knocked out by its own copy or any other particle at later time. We have chosen $a\!=\!0.03\sigma$ as this is the average distance particles traverse in unit time $\tau$. From the {\it MSD} data, we can easily identify this length scale and the corresponding time scale fall within the intermediate slow relaxation regime for all low temperature morphologies. In Fig.\ref{avdyn}(b), we show that configurations at $T\!=\!0.05$ become uncorrelated in a scale-free manner from their arbitrarily chosen initial conformations after certain residence time $\tau_r$. We will return to the dependence of $\tau_r$ with different aggregate shapes later. Here, we note that this power-law behaviour for structural relaxation changes drastically when the final temperature of the system goes above $T\!=\!0.2$. For these high temperature systems, the threshold $a\!=\!0.3\sigma$ is again the average extent of particle displacement in unit time $\tau$ and the system goes from ballistic to diffusive after this time. Considering the temperature dependence of self-overlap of configurations and {\it MSD} of constituent particles together, we infer that the underlying phase separation process affects the average dynamics of the system as the system crosses gas-liquid critical temperature during cooling. However, local dynamics is governed by thermal fluctuations set by the final constant temperature.

\subsection{Local dynamics: self and relative displacement fluctuations}
The statistics of individual displacement fluctuations for different local geometries is presented in this section. The probability of finding a normally diffusing particle is given by a Gaussian distribution with variance related to its long-time diffusivity. However, this distribution is often complemented by a {\it fat} exponential tail for systems exhibiting anomalous diffusion such as ours. We determine the probability of finding a particle at a distance ${\bf r}(t)$ from its arbitrarily chosen previous position ${\bf r}(0)$ by computing the self-part of van Hove function \cite{vH},
\begin{equation}
G_s(r,t) = 1/N\sum_{i=1}^n \langle \delta(r-|{\bf r}_i(t)-{\bf r}_i(0)|)\rangle
\end{equation}
Interestingly, a broad range of microscopic dynamics are captured by $G_s(r,t)$ for aggregates across their morphological hierarchy. At the end of intermediate slow relaxation ($t^*\!=\!50\tau$), $G_s(r,t^*)$ is dominantly Gaussian with small exponential tail (Fig.\ref{lkldyn}(a)) for compact clusters. On the contrary, Gaussian part of $G_s(r,t)$ is recessive and buried within the long exponential part (Fig.\ref{lkldyn}(b)) for non-compact aggregates. These two distinct parts immediately point to at least two different corresponding diffusivities \cite{nnGs} other than the average diffusivity and thus, the heterogeneous dynamics in the system. Most importantly, by tuning the competing interactions, we are able to capture the whole range of spatio-temporal heterogeneity for aggregates.

Next, we identify different physical mechanisms responsible for the limiting cases of such heterogeneity using the evolution of relative displacement fluctuations. This is captured by calculating the dynamic Lindemann parameter \cite{dynLnd},
\begin{equation}
\gamma(t)=\langle(\Delta{\bf u}_i(t)-\Delta{\bf u}_j(t))^2\rangle/2\sigma^2
\end{equation}
as the difference between displacement $\Delta{\bf u}(t)\!=\!{\bf r}(t)\!-\!{\bf r}(0)$ of $i$-th particle and its $j$-th neighbours. The particles in compact aggregates ($\xi\!=\!0.5,0.6$) show negligible or very little displacement with respect to their neighbours until time $\!\sim\!\tau$ (Fig.\ref{lkldyn}(c)). This leads us to conclude that the particles, in this case, are {\it caged} by their neighbours. After certain time, the particles get out of the cage and can move freely till they get into another cage and $\gamma(t)$ increases monotonically with time. However, from visual inspection, we expect this to happen more frequently for the particles at the edge of a cluster and less frequently for those at the centre of compact crystalline aggregates. A reversal of this feature is observed for non-compact aggregates. Large repulsion ($\xi\!=\!0.7,0.8$) responsible for these chain-like structures hinders the particles from coming close to each other and $\gamma(t)$ shows a sharp initial increase. Once they are able to decrease their relative distance due to thermal fluctuations, strong attraction binds them and $\gamma(t)$ shows considerable decrease in the slope. We name this feature as {\it bonding}. In essence, as the aggregate morphology changes from {\it compact to non-compact}, the dynamics of constituent particles show transition from {\it caging to bonding}.

\section{Discussion}
The mobility of particles forming aggregates becomes extremely sluggish due to the nominal thermal fluctuations present in the system at low ($T\!<\!0.2$) temperatures. The microscopic dynamics is heterogeneous following the restricted and random spatial extent of aggregates. Following the individual particle trajectories in different aggregate morphologies and the evolution of their fluctuations, we have identified the fundamental mechanisms for this continuous transition in heterogeneous dynamics. Under strong influence of short-range attraction, the movement of a particle is restricted within a cage bounded by its neighbours and particles align themselves in local hexagonal arrangement, energetically favoured in 2D. Consequently, the probability of finding a particle at any instant will be a Gaussian with variance equal to the typical cage size determined by the time window between the chosen instant and an arbitrary time origin. The probability will decay exponentially outside this typical size due to the overall rearrangements happening within the system at that instant. With increasing repulsion, the degree of geometric frustration due to competition between interactions increases and the hexagonal symmetry becomes unfavourable. An emergent anisotropy originating from the complex interplay between enhanced frustration and thermal fluctuations allows only sidewise bonding of particles. As a result, particles assume finite-size string-like non-compact structures and retain their shape as the probability of finding the constituent particles becomes exponentially smaller with increasing distance from their original position.
\begin{figure}[h!]
\begin{center}
\includegraphics[width=0.65\linewidth]{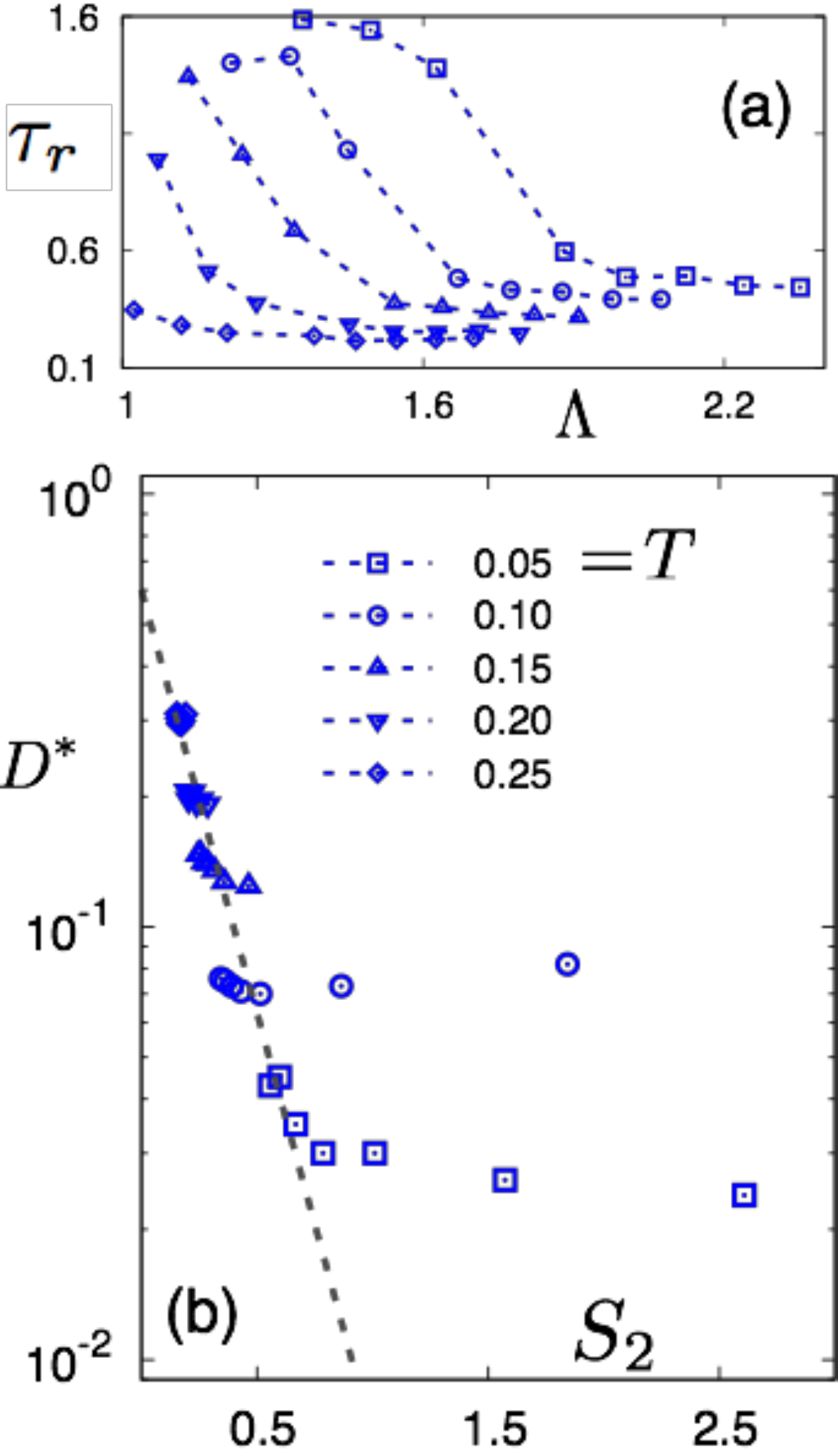}
\end{center}
\caption{
(color online) (a) Temperature dependence of reduced diffusivity $D^*$ is plotted against the dimensionless descriptor $\Lambda$ of aggregate morphologies. Please note that $\Lambda$ is also temperature dependent, it assumes lower values with increasing $T$. For low temperatures, $T\!<\!0.2$, $D^*$-$\Lambda$ relation always shows an inflection point around the structural transition. At and beyond $T=0.2$, $D^*$ becomes constant for all $\Lambda$ as the system behaves like an equilibrium liquid. (b) These observations can be understood more clearly in terms of residence time of particles $\tau_\alpha$ as a function of $\Lambda$. For example, $\tau_\alpha$ marks a clear dynamical transition with respect to $\Lambda$ at $T\!=\!0.05$. This transition goes away with increasing temperature. (c) $D^*$ can further be related to configurational randomness of the system measured by two-body excess entropy $S_2$. In fact, $D^*$-$S_2$ follows the universal relation ({\it see text}) for small values of $S_2$ or high randomness. The compact aggregates at low temperatures having local crystalline order naturally deviate from this universality.
}
\label{REStDiffS2}
\end{figure}

Following this physical picture, it is only natural to expect that particles within a crystalline compact arrangement stay longer at or in close proximity to their original position compared to those in non-compact conformations. Our computation of the self-overlap function, in fact, supports this expectation (Fig.\ref{avdyn}(b)). The long-time algebraic tail of self-overlap can be explained in terms of a {\it survival function} of the form $(\tau_r/t)^\gamma$ where $\tau_r$ is the minimum {\it residence time} of a particle and $\gamma$ is the {\it tail} or {\it Pareto index} \cite{paretolaw}. The typical {\it residence time} is, indeed, large for compact clusters and drops down at least by amount $\tau$ for non-compact conformations. Here, we mention that these typical time scales are specific to the local geometry and not to the specific values of parameter $\xi$. We recall that aggregates of statistically similar morphology can be obtained for various combinations of attraction range (fixed by $\alpha$), repulsion strength ($A$) and screening length ($\xi$). The morphological hierarchy can be classified by using a dimensionless parameter $\Lambda$ \cite{2Dmorph} which encodes all three control parameters $\{\alpha,A,\xi\}$ of the effective interaction. As $\Lambda$ faithfully describes the local geometry of aggregates, we plot $\tau_r$ corresponding to these geometries with respect to $\Lambda$ in Fig.\ref{REStDiffS2}(a). A clear transition in $\tau_r$ is observed at $\Lambda\!\sim\!2.$ which coincides with non-compact ($\Lambda\!\geq\!2.0$) to compact ($1.5\!<\!\Lambda\!<\!2.0$) transition. However, $\Lambda$ itself is a function of temperature. For higher values of final temperature (which means the cooling was stopped at higher temperature), thermal fluctuations compete with geometric frustration and finally take over at a temperature above $T\!=\!0.2$ and the system behaves like an equilibrium liquid.

Finally, we proceed further to relate the average diffusivity with the structural randomness of the system. Considering the diffusion in equilibrium simple liquids is governed by the Enskog process \cite{enskog}, it is possible to relate the diffusivity of the system to the configurational phase space volume available to the same, that is,
\begin{equation}
D=C\sigma^2\Gamma e^{S_2}
\label{eqDS2}
\end{equation}
where $\Gamma$ is the typical collision frequency and $S_2$, the two-body excess entropy \cite{exS1}, is an ensemble independent measure of structural randomness of the system \cite{exS2}. $C$ is a system dependent dimensionless non-zero constant. Such relation was phenomenologically proposed and established for simple equilibrium model systems \cite{rosenfeld,dzugutov}. Later, it was experimentally verified for colloidal systems \cite{tong} and a wide range of liquid metals \cite{hoyt}. For our model aggregates, non-compact clusters posses higher degree of randomness ($S_2\!\sim\!0$) closer to liquids. $S_2$ assumes large negative values as soon as local crystalline ordering appears within the system as in compact aggregates \cite{2Dmorph}. Since for all these morphologies, the microscopic dynamics of our model system at low temperature is non-ergodic and much slower than simple liquids, an Enskog process cannot be expected. We define a dimensionless diffusivity $D^*=D^\infty/D_0$ where $D^\infty\!=\!\lim_{t\!\to\!\infty}\langle r^2(t)\rangle$ is the so-called long time diffusivity and $D_0\!=\!\lim_{t\!\to\!0}\langle r^2(t)\rangle$ represents particle motion in short-time ballistic motion when the particles have yet to experience the presence of their neighbours. Plotting $D^*$ against $S_2$ (Fig.\ref{REStDiffS2}(b)), an exponential relation, $D^*\!\sim\!e^{S_2}$ (similar to eq.\ref{eqDS2}) is observed for different temperatures and small values of $S_2$. We note that small $S_2 (\le\!1)$ refers to equilibrium liquids for $T\!>\!0.2$ and non-equilibrium non-compact structures for $T\!<\!0.2$. This observation is surprising and implies that the Enskog process is not limited to equilibrium but related to the morphological randomness. As the system adapts specific crystalline symmetry this feature breaks down and $D^*$ becomes nearly independent of $S_2$ for the compact aggregates. With the possibility of a fundamental description of microscopic dynamics in terms of the degree of configurational randomness, this observation surely requires further study.



To conclude, aggregates are suitable candidates to relate non-ergodic dynamics to the corresponding local morphology. Geometric frustration inherent to the competing short-range attraction and long-range repulsion leads to non-ergodic stationary aggregates. The average dynamics bearing the effects of phase separation at sufficiently low temperature shows long time sub-diffusive behaviour following intermediate slow relaxation. Both structural randomness of aggregates and the heterogeneous micro-dynamics of constituent particles can be changed seamlessly by tuning the interactions for fixed thermodynamic conditions. Specifically, the particles in compact aggregates experience caging from their nearest neighbours and individual dynamics are close to that of particles in glassy systems. On the other hand, bonding of particles is prevalent in non-compact clusters and the individual dynamics resembles that of particles moving in a network. Further correspondence between dynamics and morphology of aggregates has been established by relating diffusivity and two-body excess entropy. In future, we plan to investigate the effect of cooling rate, ageing time and external mechanical fields on these pattern forming systems to understand the role of local geometry in systems with unusual dynamical responses.

\acknowledgments
TD and MMB were supported by the OIST Graduate University with subsidy funding from the Cabinet Office, Government of Japan. TL conducted this work under the auspices of the National Nuclear Security Administration of the U. S. Department of Energy at Los Alamos National Laboratory under Contract No. DE-AC52-06NA25396. 


\begin{thebibliography}{}
\bibitem{effort1} M. D. Ediger, {\it Annu. Rev. Phys. Chem.} {\bf 51}, 99 (2000).
\bibitem{effort2} E. R. Weeks, J. C. Crocker, A. C. Levitt, A. Schofield, and D. A. Weitz, {\it Science} {\bf 287}, 627 (2000).
\bibitem{effort3} L. Berthier and G. Biroli, {\it Rev. Mod. Phys.} {\bf 83}, 587 (2011).
\bibitem{effort4} P. J. Lu and D. Weitz, {\it Annu. Rev. Condens. Matter Phys.} {\bf 4}, 217 (2013).
\bibitem{DH1} P. G. Wolynes, V. Lubchenko, {\it Structural Glasses and Supercooled Liquids: Theory, Experiment, and Applications} John Wiley \& Sons, (2012).
\bibitem{DH2} L. Berthier, G. Biroli, J.-P. Bouchaud, L. Cipelletti, W. van Saarloos {\em ed.}, {\it Dynamical Heterogeneities in Glasses, Colloids, and Granular Media} Oxford University Press, (2011).
\bibitem{conf1} J. R. Bordin, A. B. de Oliveira, A. Diehl and M. C. Barbosa, {\it J. Chem. Phys.} {\bf 137}, 084504 (2012).
\bibitem{conf2} L. B. Krott and M. C. Barbosa, {\it J. Chem. Phys.} {\bf 138}, 084505 (2013).
\bibitem{cell} D. Holcman and Z. Schuss, {\it J. Phys. A: Math. Theor.} {\bf 47}, 173001 (2014).
\bibitem{aggrgt1} J. Groenewold and W. K. Kegel, {\it J. Phys. Chem. B} {\bf 105}, 11702 (2001).
\bibitem{aggrgt2} G. Malescio and G. Pellicane, {\it Nat. Mater.} {\bf 2}, 97 (2003).
\bibitem{aggrgt3} F. Sciortino, S. Mossa, E. Zaccarelli, and P. Tartaglia, {\it Phys. Rev. Lett.} {\bf 93}, 055701 (2004).
\bibitem{expt1} A. Stradner, H. Sedgwick, F. Cardinaux, W. C. K. Poon, S. U. Egelhaaf, and P. Schurtenberger, {\it Nature} {\bf 432}, 492 (2004).
\bibitem{expt2} T. H. Zhang, J. Klok, R. H. Tromp, J. Groenewolda, and W. K. Kegel, {\it Soft Matter} {\bf 8}, 667 (2012).
\bibitem{ps1} M. Carpineti, and M. Giglio, {\it Phys. Rev. Lett.} {\bf 86} 3327 (1992).
\bibitem{ps2} M. Muschol, and F. Rosenberger, {\it J. Chem. Phys.} {\bf 107}, 1953 (1997).
\bibitem{arrest} P. Segre, V. Prasad, A. Schofield, and D. Weitz, {\it Phys. Rev. Lett.} {\bf 86}, 6042 (2001).
\bibitem{neqhet1} A. M. Kulkarni, N. M. Dixit, and C. F. Zukoski, {\it Faraday Discuss.} {\bf 123}, 3 (2003).
\bibitem{neqhet2} A. M. Puertas, M. Fuchs, and M. E. Cates, {\it J. Chem. Phys.} {\bf 121}, 2813 (2004).
\bibitem{dbt1} D. A. Weitz, and M. Oliveria {\it Phys. Rev. Lett.} {\bf 52} 1433 (1984).
\bibitem{dbt2} D. A. Weitz, J. S. Huang, M. Y. Lin, J. Sung {\it Phys. Rev. Lett.} {\bf 54} 1416 (1985).
\bibitem{dbt3} G. Foffi, C. De Michele, F. Sciortino, and P. Tartaglia {\it Phys. Rev. Lett.} {\bf 94} 078301 (2005).
\bibitem{dbt4} B. Ruzicka, E. Zaccarelli, L. Zulian, R. Angelini, M. Sztucki, A. Moussa•d, T. Narayanan, and F. Sciortino, {\it Nature materials} {\bf 10}, 56 (2011).
\bibitem{dbt5} E. Zaccarelli, P. J. Lu, F. Ciulla, D. A. Weitz, and F. Sciortino {\it J. Phys. Condens. Matter} {\bf 20} 494242 (2008).
\bibitem{dbt6} J. C. F. Toledano, F. Sciortino, E. Zaccarelli {\it Soft Matter} {\bf 5}, 2390 (2009).
\bibitem{dbt7} F. Cardinaux, E. Zaccarelli, A. Stradner, S. Bucciarelli, B. Farago, S. U Egelhaaf, F. Sciortino, and P. Schurtenberger {\it J. Phys. Chem. B} {\bf 115}, 7227 (2011)
\bibitem{glss1} E. Zaccarelli, G. Foffi, K. A. Dawson, S. V. Buldyrev, F. Sciortino, and P. Tartaglia, {\it Phys. Rev. E} {\bf 66}, 041402 (2002).
\bibitem{glss2} N. Gnan, G. Das, M. Sperl, F. Sciortino, and E. Zaccarelli, {\it Phys. Rev. Lett.} {\bf 113}, 258302 (2014).
\bibitem{lebo_pen} J. L. Lebowitz, O. Penrose, {\it J. Math. Phys.} {\bf 7}, 98 (1966);
\bibitem{compint} J. L. Lebowitz, {\it Ann. Rev. Phys. Chem.} {\bf 19}, 389 (1968).
\bibitem{colloid1} P. J. Lu, E. Zaccarelli, F. Ciulla, A. B. Schofield, F. Sciortino and D. A. Weitz, {\it Nature}, {\bf 453}, 499 (2008).
\bibitem{protein1} R.P. Sear {\it J. Chem. Phys.} {\bf 111}(10) 4800 (1999).
\bibitem{protein2} A. Shukla, E. Mylonas, E. Di Cola, S. Finet, P. Timmins, T. Narayanan, and D. I. Svergun, {\it Proc. Natl. Acad. Sci.} {\bf 105}, 5075 (2008).
\bibitem{dna_np} S. Srivastava, D. Nykypanchuk, M. Fukuto, J. D. Halverson, A. V. Tkachenko, K. G. Yager, and O. Gang, {\it J. Am. Chem. Soc.} {\bf 136}, 8323 (2014).
\bibitem{salt} C. N. Likos, {\it Physics Reports} {\bf 348} 267 (2001).
\bibitem{globprot} G. A. Vliegenthart and H. N. W. Lekkerkerker, {\it J.Chem.Phys.} {\bf 112} (12) 5364 (2000).
\bibitem{colloid2} A. Puertas, C. De Michele, F. Sciortino, P. Tartaglia and E. Zaccarelli {\it J. Chem. Phys.} {\bf 127} 144906 (2007).
\bibitem{2n-n} G. A. Vliegenthart, J. M. F. Lodge, and H. N. W. Lekkerkerker, {\it Physica A} {\bf 263} 378 (1999).
\bibitem{2Dmorph} T. Das, T. Lookman, and M. M. Bandi, \texttt{arXiv:1502.03161}
\bibitem{lammps} Available at \texttt{http://lammps.sandia.gov}
\bibitem{langevin} T. Schneider and E. Stoll, {\it Phys. Rev. B} {\bf 17}, 1302 (1978).
\bibitem{critical} P. Charbonneau and D.R. Reichman, {\it Phys. Rev. E} {\bf 75}(5) 050401 (2007).
\bibitem{nergo1} D. Thirumalai, R. D. Mountain, and T. R. Kirkpatrick, {\it Phys. Rev. A} {\bf 39}, 3563 (1989).
\bibitem{nergo2} R. D. Mountain and D. Thirumalai, {\it J. Phys. Chem.} {\bf 93}, 6975 (1989).
\bibitem{msd} S. Havlin and D. Ben-Avraham, {\it Advances in Physics} {\bf 51} 187 (2002).
\bibitem{sfd} S. Herrera-Velarde, A. Zamudio-Ojeda, and R. Castaneda-Priego {\it J. Chem. Phys.} {\bf 133}, 114902 (2010).
\bibitem{ovrlp1} G. Parisi, {\it J. Phys. A: Math. Gen.} {\bf 30}, L765 (1997).
\bibitem{ovrlp2} S. C. Glotzer, V. N. Novikov, and T. B. Schroder {\it J. Chem Phys.} {\bf 112}, 509 (200)
\bibitem{vH} L. van Hove, {\it Phys. Rev.} {bf 95} 249 (1954).
\bibitem{nnGs} B. Wang, J. Kuo, S. C. Bae and S. Granick, {\it Nature Materials} {\bf 11}, 483 (2012).
\bibitem{dynLnd} K. Zahn, R. Lenke, and G. Maret, {\it Phys. Rev. Lett.} {\bf 82} 2721 (1999).
\bibitem{paretolaw} M. E. J. Newman, {\it Contemporary Physics} {\bf 46}, 323?351 (2005).
\bibitem{enskog} S. Chapman and T.G. Cowling, The Mathematical Theory of Non-Uniform Gases (Cambridge University Press, London, 1970).
\bibitem{exS1}R .E. Nettleton and M. S. Green, {\it J. Chem. Phys.} {\bf 29} 1365 (1958).
\bibitem{exS2} A. Baranyai and D.J. Evans, {\it Phys. Rev. A} {\bf 40} 3817 (1989).
\bibitem{rosenfeld} Y. Rosenfeld, {\it Phys. Rev. A} {\bf 15}, 2545-2549 (1977).
\bibitem{dzugutov} M. Dzugutov, {\it Nature} {\bf 381}, 137 (1996).
\bibitem{tong} X. Ma, W. Chen, Z. Wang, Y. Peng, Y. Han, and P. Tong, {\it Phys. Rev. Lett.} {\bf 110}, 078302 (2013).
\bibitem{hoyt} J. Hoyt, M. Asta, and B. Sadigh, {\it Phys. Rev. Lett.} {\bf 85}, 594-597 (2000).
\end{thebibliography}
\end{document}